# What Cost Knowledge Management? The Example of Infosys

Chris Kimble

**Abstract**
The term knowledge management (KM) first came to prominence in the late 1990s. Although initially dismissed as a fad, KM continues to be featured in articles concerning business productivity and innovation. And yet, clear-cut examples that demonstrate the success of KM are few and far between. A brief examination of the history of KM explores the reasons for this and looks at some of the assumptions about what KM can achieve. A subsequent analysis of the experiences of Infosys with KM shows that for KM to be successful, organizational leaders need to engage in a continuous process of modification and maintenance. Although KM initiatives can be made to yield worthwhile returns over an extended period, there are often substantial ongoing costs associated with them.

## 1   Introduction

Many saw the years of economic growth between the recession of the late 1980s and the bursting of the dot-com bubble in early 2000 as "the beginnings of a great new era, a third industrial revolution, founded on new technologies rooted in computers and the potential of new information technologies" (Harris, 2001, p. 22).

In the world of business and economics, new theories were emerging that talked about the growth of a knowledge-based economy, which stressed the importance of knowledge as a source of sustainable competitive advantage (Stewart et al., 1991). For example, Nonaka claimed that knowledge had become "the one true source of lasting competitive advantage" (Nonaka, 1991, p. 96). The growth of the resource-based view of the firm (Grant, 1996) contributed to the idea that knowledge might be a resource that could be managed in the same way as other, more traditional resources.

Building on earlier work, such as Pentland (1995), authors such as O'Leary (1998) began to argue that the future prosperity of business was tied to the development of IT-based knowledge management systems (KMSs) that could capture and manage a firm's knowledge assets. Yet, even though $73 billion was being spent on software related to knowledge management (KM) (Murphy & Hackbush, 2007) and $8 billion was being spent on KM-related services (Gupta, Sharma, & Hsu, 2008), unequivocal and well documented examples of successful KM projects, such as those that appeared in the early days of KM (Davenport, DeLong, & Beers, 1998), remain few and far between. What could be the reason for this?

What follows is a brief review of the history of KM, a look at some of the assumptions upon which it is founded, an assessment of the effect that these assumptions might have on the expected return on investment, and an analysis of a case study of an organization that has successfully implemented KM: Infosys, a business technology consulting and IT services company. Infosys has successfully implemented KM over a period of several years; the case study presented here is based on accounts published in books, reports, and peer-reviewed academic journals. It demonstrates the way in which Infosys has had to continually monitor and review its approach to KM to ensure that it continues to provide value for its businesses. It also shows how successful KM initiatives consist of a delicate balancing act between top-down initiatives that capture and codify knowledge and allowing scope for groups and communities to find their own solutions.

## 2 A Brief History of Knowledge Management

The different approaches that have been applied to KM are usually described in terms of generations (McElroy, 2002). First-generation KM is seen as being focused on capturing knowledge and storing it in IT-based repositories, whereas second-generation KM has a stronger focus on people and communities as sources of knowledge. In the sections that follow, we will look at these two viewpoints in more detail and explore some of the ideas that underpin them.

### 2.1 First-Generation Knowledge Management

At least initially, technology appeared to be the obvious solution to the growing need to manage knowledge and, in the early days, KM was dominated by the idea that knowledge could be captured and stored for later use in some form of IT-based repository. The reasons for this are not difficult to understand.

During the early 1990s there had been steady progress with representing and modeling knowledge in the field of computer science and, by the late 1990s, authors such as Tsui, Garner, and Staab (2000) were arguing strongly for the role of artificial intelligence in KM. Similarly, in the information systems field, work on IT-based organizational memories (Stein & Zwass, 1995) appeared to offer a promising approach to realizing IT-based KM. Much of what was being written about KM at the time was concerned with the promise of increasing returns (Teece, 1998) from the creation of databases of expertise (Davenport et al., 1998). These, it was claimed, would make the "treasure house" of knowledge that existed inside organizations more widely available (O'Dell & Grayson, 1998).

Much of the thinking that underpinned these beliefs was based on Shannon and Weaver's (1949) earlier work on communications theory which, although 40 years old at the time, was still influential (Zander & Kogut, 1996). Shannon and Weaver's work was concerned with how information could be turned into an encoded message and transmitted via a communication network. They showed that, as long as a suitable "codebook" for encoding and decoding the message existed, a message could be transmitted accurately. Their theory had revolutionized communications and the belief at the time was that if knowledge could be treated in the same way - that is, if it could be "codified" - it, too, could be turned into information (Zack, 1999).

Exhibit 1. Assumptions About the Growth of Knowledge Made in First-Generation KM

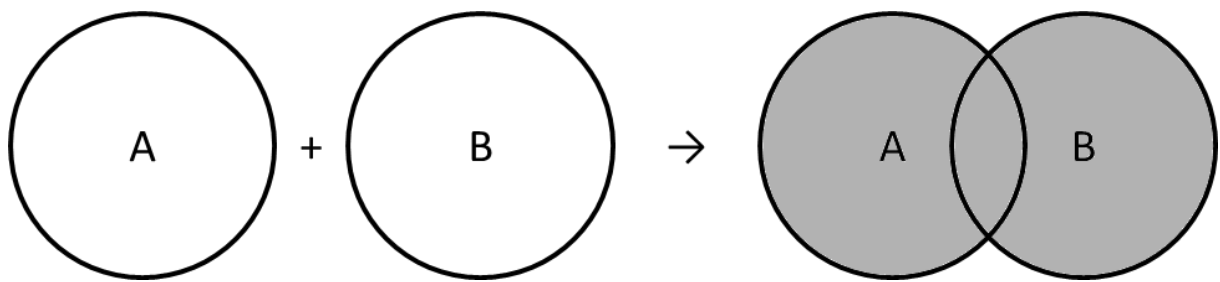

Codified knowledge has some attractive properties. It has many of the features of a standard commodity, which means, for example, that it can be bought and sold. In addition, because it is based on a code that can be understood anywhere, the cost of its transmission and reproduction is extremely low. As a commodity, codified knowledge also has some unusual features, such as the "stock" of available knowledge not being reduced by its "sale." However, perhaps its most important feature is that as new knowledge is discovered, it can simply be added to the existing stock of available knowledge, which will then grow cumulatively. This process can be visualized as a union of two sets (see Exhibit 1).

According to Cohendet and Steinmueller (2000), this form of growth should lead to a "leveraging" effect that increases return on investment over time.

Although the view of increasing returns from knowledge codification was widespread at the time, there were some dissenting voices. For example, McDermott (1999) argued that IT alone could not bring about the "leveraging" of knowledge, as it did not take into account the human dimension of KM. The doubts of those such as McDermott concerning the viability of this approach, combined with the disappointing results from many first-generation KM projects, led to the growth of an alternative, second-generation approach to KM.

### 2.2   Second-Generation Knowledge Management
The key to understanding the difference in approach between first and second-generation KM revolves around this distinction between tacit and explicit knowledge: "'Explicit' or codified knowledge refers to knowledge that is transmittable in formal, systematic language. On the other hand, 'tacit' knowledge has a personal quality, which makes it hard to formalize and communicate" (Nonaka, 1994, p. 16)

In first-generation KM, the goal had been to make knowledge explicit so that it could easily be transmitted throughout an entire organization. Such knowledge, however, is, by its nature, "leaky" and prone to be copied (Liebeskind, 1996). Therefore, tacit knowledge, which was seen as being "sticky" and difficult to copy (von Hippel, 1994), became the focus of interest. For example, Grant (1996, p. 376) concluded that sustainable competitive advantage "requires resources which are idiosyncratic (and therefore scarce), and not easily transferable or replicable. These criteria point to knowledge (tacit knowledge in particular) as the most strategically important resource which firms possess."

The key to being able to treat knowledge as though it were information is the establishment of a shared "codebook." This does not present many problems when dealing with stable physical entities, such as communications systems; however, the same cannot be said of tacit knowledge. The laws of physics are the same everywhere; however, tacit knowledge is embedded in people and their interactions with the world around them. This means that as people learn and adapt, their tacit knowledge will change. Similarly, because tacit knowledge is based on personal experience, it will be different from one location to another. Furthermore, not only does the knowledge itself change, but, as people develop their own "shorthand" to describe what they do, the language used to describe it also will change.

The solution to the problem of dealing with this type of knowledge was to shift the focus of attention from individuals to communities. Although not all members of a community are the same, the fact that they are in the same community implies that they share a set of common beliefs. The argument is that this shared worldview will, in effect, act as a surrogate for the codebook found in the first-generation approach to KM and, thus, facilitate the sharing of knowledge. Communities of practice (Wenger, McDermott, & Snyder, 2002) - groups created through a process of socialization around some communal activity - became a particular focus of interest for this approach to KM, as such groups not only create the basis for a common understanding, but also create their own language that can be used to express this understanding to others.

Although this approach provides the solution to one set of problems, it creates others. For example, a community's "private" language, which eases the sharing of knowledge among its members, can act as a barrier between it and the world outside (Kimble, Grenier, & Goglio-Primard, 2010). The fact that the sharing of knowledge depends on the existence of a shared worldview also makes the assumptions of increasing returns over time harder to justify. The benefits of attempts to manage knowledge now come from establishing "cognitive common ground" between individuals; in this case, the growth of the "stock" of available knowledge more closely resembles the intersection of two sets than the more optimistic union of sets in first-generation KM (see Exhibit 2). According to Cohendet and Steinmueller (2000), the best that can be expected from this scenario is a flat rate of return.

Exhibit 2. Assumptions About the Growth of Knowledge Made in Second-Generation KM

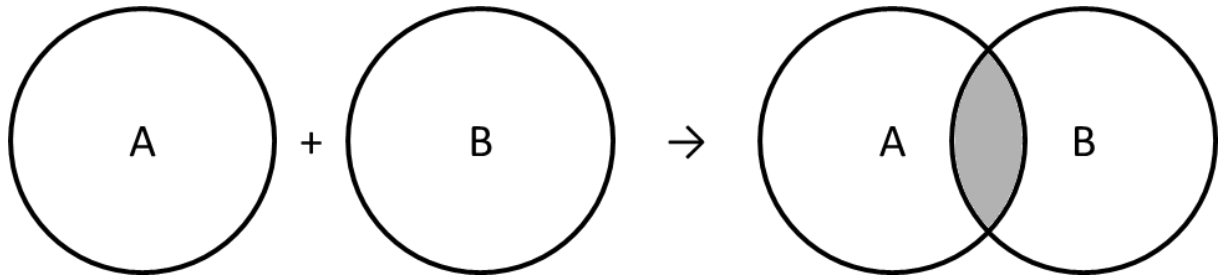

## 3   Infosys: A Success Story for Knowledge Management?

As a company, Infosys is a remarkable success story. It was founded in Pune, India, in 1981 by seven entrepreneurs with a total working capital of 10,000 rupees (approximately $180 in today's economy). At the time, it took approximately three years to get the license needed to import a computer into India and two years to install a telephone line. When the company was founded, it had neither (Mishra & Chandran, 2011). Infosys now employs almost 154,000 people and has a presence in 29 countries. According to the most recent listing on NASDAQ, in the year ending March 2012, the company had annual revenue of almost $7 billion and made an operating profit of $1.7 billion. Much of this success has been attributed to the company's attitude regarding the management of its intellectual capital: "Our vision is to make every instance of learning within Infosys available to every employee. We want the collective learning of Infosys to back every person wherever they might be located within the company. Our motto is 'Learn once, use anywhere'" (Garud, Kumaraswamy, & Sambamurthy, 2006b, p. 224).

The company's approach is generally in line with the first-generation approach to knowledge management outlined earlier: the vision is of knowledge being captured, stored, and leveraged to provide ever-increasing returns. The principal method used to achieve this is codification. That is, each time a new item of knowledge is identified, it is analyzed, documented, stored, and cross-indexed in some way so that it is available for later retrieval and reuse. This process, however, is not always as straightforward as it may seem and, in practice, requires constant monitoring and adjustment.

Infosys's interest in KM began as early as 1985 (Garud et al., 2006b, p. 225), but the first significant KM initiative at the company did not occur until 1992, with the creation of bodies of knowledge (BOKs). These were paper-based accounts that were written by employees about their on-the-job experiences and distributed throughout the company. As the technology behind the World Wide Web began to develop, these paper-based BOKs began to gradually migrate toward more easily distributed HTML documents, and, in 1996, Infosys launched Sparsh (meaning "in touch"), a company-wide Intranet that facilitated the linking of local repositories and simplified the distribution of the HTML-based BOKs.

The company was growing fast; much of its work was based on relatively short-term project work, and Infosys groups and teams were beginning to find it difficult to locate the internal experts they needed. In 1997, Infosys created a "yellow pages" directory on Sparsh, called the People-Knowledge Map, which listed the names and contact details of experts in specific fields in the company (Garud, Kumaraswamy, & Sambamurthy, 2006a). This was a voluntary scheme that allowed various internal experts to "go public" so that they could be more easily consulted by colleagues who needed advice. Developments such as this and domain-specific mailing lists and bulletin boards soon lead to Sparsh becoming the host to a range of virtual groups and social networks, which created a "campus-like environment" within the company (Garud et al., 2006b, p. 19).

Although the idea of learnability (the recruitment of an intellectually flexible and adventurous workforce with a passion for learning) and that of sharing and reusing knowledge had always been part of the Infosys philosophy, until this point the company had never needed to formalize its KM efforts. But it had now grown in size and geographical spread. At the turn of the century, it faced a major challenge. It saw that work related to the Y2K problem (the name given to dealing with systems whose internal calendar did not go past December 31, 1999) would soon decline, while another market for e-commerce - an area in which it had little expertise - was beginning to grow rapidly. It needed to find a way to manage a smooth transition from being a business driven by dealing with the Y2K problem to one driven by e-commerce.

In line with the previous approach of the company, the answer was seen as KM. There was considerable debate over which model of KM should be followed: the centralized, IT based, first-generation model or the more widely distributed, community based approach of second-generation KM (Garud et al., 2006a). In late 1999, a decision was reached: The company created a new internal group to manage KM and launched a coordinated, company-wide KM program based around KShop, a knowledge portal that would bring together, in one centralized repository, all the knowledge that resided in different groups spread throughout the company.

In accordance with its previous approach to KM, KShop was broadly consistent with the first-generation view of KM. It was described as "centrally facilitated yet organizationally distributed" (Garud et al., 2006b, p. 225), providing an architecture that allowed it to be deployed throughout the firm while allowing sufficient flexibility for business units to take ownership of specific content areas. KShop was organized into hierarchal categories of "knowledge assets." Before any new knowledge asset could be added to the repository, a pool of volunteer reviewers would screen it to ensure its quality and to identify any potential issues related to intellectual property. It contained a number of different sections, such as BOKs, case studies, and reusable artifacts, and offered employees a range of tools that they could use to create and manage their own web pages.

Despite the importance of KShop, by the end of 2000, the system appeared to be running into problems. Less than 5 percent of employees had made any form of contribution, and even fewer had used the information the system contained. Most teams and communities simply continued to use their existing networks, thus preventing the companywide sharing of knowledge that had been intended. The response of the KM group was to introduce the Knowledge Currency Unit (KCU). KCUs could be exchanged for monetary rewards or gifts and were awarded to employees who made or reviewed contributions to KShop. In addition, a league table of contributors to KShop was implemented to give personal recognition to star contributors. Within a year, more than 130,000 KCUs had been distributed, more than 2,400 new knowledge assets had been registered, and almost 20 percent of Infosys's employees had made at least one contribution (Garud & Kumaraswamy, 2005).

The pendulum at Infosys had now swung the other way. The KM group began to receive reports that KShop contained too many items that were out of date or of little use, and the people whose task it was to review new additions complained that they were overwhelmed. Finally, when new additions of questionable quality began to receive high-quality ratings from the colleagues of those who submitted them, the entire scheme was put under scrutiny. It was found that people who had previously contributed simply "for the joy of sharing" (Garud & Kumaraswamy, 2005, p. 22) were becoming increasingly motivated by monetary rewards. Additionally, and perhaps more seriously, there were also signs that certain groups were turning away from KShop altogether and reverting to building and maintaining their own "clean" local repositories, thus undermining the whole rationale for the introduction of KShop.

In April 2002, the entire KCU scheme was revised. The value of the monetary rewards was reduced and the level at which KCUs could be exchanged for cash benefits was raised; greater emphasis was given to feedback from the end-users, as opposed to that of colleagues; and corroborative evidence was now required before a knowledge asset could be given a very high rating. Within six months, the number of new contributors had fallen by 37 percent and the number of new knowledge assets fell by 26 percent (Garud & Kumaraswamy, 2005). After this initial decline, the number of contributors and contributions began to flatten out and, by 2003, KShop had stabilized and was described as a fully functioning knowledge portal (Garud et al., 2006a).

As KShop grew, however, the number of categories it contained also began to multiply. It soon became clear that no single categorization scheme could serve all the different communities in Infosys, each of which had its own specialized language. The solution to this new problem was to create a "system imposed demand" (Garud et al., 2006b, p. 225). The goal of this approach was that knowledge capture and codification would be seamlessly incorporated into a group's everyday activities. Project templates were modified and a new project-tracking tool implemented that automatically captured details of what went on in projects as the project teams went about their routine tasks. This approach allowed teams to classify knowledge assets in ways that were meaningful to their project while simultaneously providing KShop with the data necessary to create an organization-wide taxonomy described as "not just a framework for classifying content [but] a strategy to unify multiple constituencies" (Garud et al., 2006a, p. 282).

Despite all these efforts, levels of knowledge codification continued to vary across teams. To combat this, various forms of second-generation KM techniques were deployed. "Knowledge brokers" were introduced to encourage the sharing of knowledge and best practices across teams. Each project team nominated a "KM prime" who was responsible for identifying and facilitating the fulfillment of the team's internal knowledge needs during a project and who, after its completion, took responsibility for codifying and sharing any new knowledge gained with others. In addition, the role of "knowledge champion" was created to facilitate the organization-wide distribution of knowledge when new technologies or techniques of particular importance were identified (Garud & Kumaraswamy, 2005).

Infosys has continually adapted its approach to meet changing circumstances. Presently, the company offers a range of services, such as cloud computing, infrastructure management, and outsourcing, which involve managing and coordinating the activities of a geographically distributed workforce. The KM challenges it now faces are those associated with virtual teams: managing knowledge across different physical locations, business units, and time zones (Kimble, 2011). Despite the potentially pejorative "first-generation" label, the approach taken by Infosys continues to rely heavily on centralized storage. According to Kurhekar and Ghosal (2009, p. 34), "for a virtual team, easy access to the knowledge repository is the key factor for the success of KM."

As Garud and Kumaraswamy (2005) point out, harnessing knowledge to produce increasing returns - the hallmark of the vision of the first-generation approach to KM - is not an easy task. Infosys provides one of the few examples of a company that has been able to sustain this approach to KM over a period of time. Following are some of the lessons that can be learned from its experience.

## 4   What Cost KM?

With the benefit of hindsight, it is easy to see that the early literature on KM was overly optimistic about what could be achieved by the capture-codify-store approach to KM. A number of articles that are highly critical of this approach have since been written (Duguid, 2005; Walsham, 2001). Nevertheless, there are a small number of examples of this approach being used successfully, although achieving that success appears to require a sustained managerial effort and an ongoing investment. Clearly, the spectacular and continued success of Infosys indicates that the first-generation label should not necessarily be taken to mean that this approach has been superseded or is out of date. It is equally clear, however, that this is not an easy path to follow and one that contains a number of potential pitfalls.

First, the issue of codification itself raises several issues. As explained previously, the ability to encode and store something in one context so that it can be decoded and used in another is a critical aspect of KM. Attempting to codify aspects of tacit knowledge is problematic, both because of the culture and because of location-specific nature of such knowledge and because of its inherently limited lifespan. Dealing with the context in which the knowledge will be used is a major problem with KM. One case study of the UK Post Office's attempt to capture and codify knowledge from the Argentinian Post Office (Hall, 2006) for use in the United Kingdom demonstrates the difficulties of achieving this. Similarly, the importance of "organizational forgetting" - that is, unlearning lessons that are no longer relevant to an organization - has long been recognized (Bowker, 1997). When knowledge is encoded and stored in a durable form, such as an IT-based knowledge repository, however, this adds an extra dimension to the problem of managing organizational memory.

Second, there is the now well-recognized balancing act between capturing knowledge as it is created and destroying the processes that are responsible for its creation (Brown & Duguid, 2000). In the early days of Eureka, a KM project concerned with improving the efficiency of repairs made to photocopiers, Xerox attempted to override the informal, practical knowledge that came from the people who repaired the machines, the "tech reps," in favor of the formalized technical knowledge provided by the engineers who were responsible for their design (Brown & Duguid, 1991). Xerox ran into similar problems to Infosys's experience with knowledge currency units when a later version of the Eureka system, which had worked well on a voluntary basis in France, was deployed in Canada, where financial incentives were given to encourage the contribution of "tech-tips" to the system (Bobrow & Whalen, 2002). KM initiatives must inevitably involve a balance between leaving room for human creativity and the constraints required to make technological solutions viable.

Finally, what is the cost of knowledge management? Economists like Cowan, David, and Foray (2000) have studied the question of the cost of codifying knowledge and conclude that the costs can vary widely, depending on how well formalized the body of knowledge that needs to be encoded already is. Others argue that the greatest returns from this approach come from the initial efforts at codification - for example, the relatively recent interest in KM in the construction industry (Dave & Koskela, 2009) - after which returns inevitably decline. Others highlight the strategic risks of codifying knowledge, pointing to the risk of competency traps and the loss of creativity (Bettiol, Maria, & Grandinetti, 2012).

The experience of Infosys clearly shows that if a knowledge management initiative is to provide a continuing return, it will require an unremitting managerial effort to ensure that it is functioning in the way that it supposed to, and a continual investment in the knowledge repositories and the maintenance of the "codebooks" needed to access them. These ongoing costs were not predicted in the original vision of first-generation KM, which foresaw increasing returns from a stock of knowledge that continued to grow as new knowledge was added to it. Ten years later, expectations are more realistic and, to paraphrase a quote often attributed to Thomas Jefferson, we now know that the real price of effective and continuing KM is eternal vigilance.